# A Precoding Aided Space Domain Index Modulation Scheme for Downlink Multiuser MIMO Systems

Nuno Souto, *Senior Member, IEEE*, and Américo Correia, *Senior Member, IEEE*

*Abstract*—In this correspondence, we propose a space domain index modulation (IM) scheme for the downlink of multiuser multiple-input multiple-output (MU-MIMO) systems. Instead of the most common approach where spatial bits select active receiver antennas, in the presented scheme the spatial information is mapped onto the transmitter side. This allows IM to better exploit large dimensional antenna settings which are typically easier to deploy at the base station. In order to mitigate inter-user interference and allow single user detection, a precoder is adopted at the BS. Furthermore, two alternative enhanced signal construction methods are proposed for minimizing the transmitted power or enable an implementation with a reduced number of RF chains. Simulation results for different scenarios show that the proposed approach can be an attractive alternative to conventional precoded MU-MIMO.

## I. INTRODUCTION

The ever-growing demand for faster and more energy efficient communications has been driving research towards finding newer and more effective solutions for the physical layer of beyond 5G (B5G) networks. Amongst the different possible approaches, index modulation (IM) schemes [1] have been captivating a lot of research interest in the wireless community. IM schemes rely on the indices of the building blocks to convey additional information bits and are able to offer interesting trade-offs in terms of error performance, complexity, and spectral efficiency (SE), making them potential candidates for B5G [2]. Well-known examples of IM are spatial modulations (SM) [3] and generalized SMs (GSM) [4], which are multiple-input multiple-output (MIMO) schemes where part of the information bits are used for selecting one or more active antennas, which then transmit $M$-ary modulated symbols. A lot of research work has been done on SM/GSM, mostly focusing on single-user (SU) [5]-[7], with only a few works also addressing uplink [8]-[10] and downlink [11]-[19] multiuser (MU) scenarios. In the case of downlink, most of the proposed space domain IM schemes typically fit into two main types of approaches. The first one [11]-[14] is often referred to as receive SM (R-SM), and is based on the idea of using part of the information bits for selecting a subset of receiver antennas that will receive the intended signals free of inter-channel interference. R-SM resorts to a specially designed precoder or to several preprocessing steps that can reduce the complexity of the receivers. They can also be designed to reduce training overhead and uplink channel state information at the transmitter (CSIT) feedback, as in [14]. Even though only a small subset of the antennas receives symbols, R-SM schemes require that they always remain active. Furthermore, the small number of antennas that can typically be deployed at the mobile equipment limits the SE gains. The second approach [15]-[19] avoids this limitation by mapping the spatial bits to the transmitter side. In [15] the authors proposed a precoder that allows the MU system to be decomposed into independent SU-SM systems. Each user's bits are mapped to an active transmit antenna index that transmits an $M$-QAM symbol. At the receiver, single stream maximum likelihood (ML) detection is applied. While only SM is considered, it can be directly extended to GSM. In [16], the authors assume single antenna receivers in a MU-SM system and also design a precoder which eliminates MU interference (MUI) enabling SU ML detection. This approach was extended in [17] to a reduced GSM version where the same amplitude-phase modulation (APM) symbol is sent in all the active antennas. While this scheme increases the SE of MU-SM [16], it falls short from the GSM version from [4]. In [18], block diagonalization (BD) precoding is combined with GSM. Spatial bits are broadcast to all users and only the conventional modulated symbols carry unicast messages. BD is applied to remove MUI while a ML detector searches amongst all the different possibilities of active antenna combinations. In [19] a linear precoding scheme that eliminates MUI and allows a simplified detection process was presented. It is designed for the same reduced GSM version as [17] and, thus, suffers from the same SE limitation. Furthermore, precoding is applied to the active antennas only, which can make the detection very challenging.

Motivated by the work above, in this correspondence we consider the downlink of a MU-MIMO system where a base station (BS) transmits precoded space domain IM symbols. To improve the SE of the schemes in [15]-[19], the approach is based on a virtual GSM applied at the transmitter side where different $M$-ary modulated symbols are sent on virtual active antennas. A precoder is applied in order to mitigate MUI and allow SU GSM detection. To enable an efficient implementation of the system, two alternative enhanced signal construction methods are presented. The main contributions of this letter can be summarized as follows:

- A space domain IM scheme that we refer to as precoding-aided transmitter side space domain index modulation (PTSDIM) is designed for the downlink of a MU-MIMO system. The method uses a precoder applied to the whole set of antennas (active and inactive), which mitigates MUI

N. Souto and A. Correia are with the ISCTE-University Institute of Lisbon and Instituto de Telecomunicações, 1649-026 Lisboa, Portugal (e-mail: nuno.souto@lx.it.pt, americo.correia@iscte-iul.pt).

This work is funded by FCT/MCTES through national funds and when applicable co-funded by EU funds under the project UIDB/EEA/50008/2020.

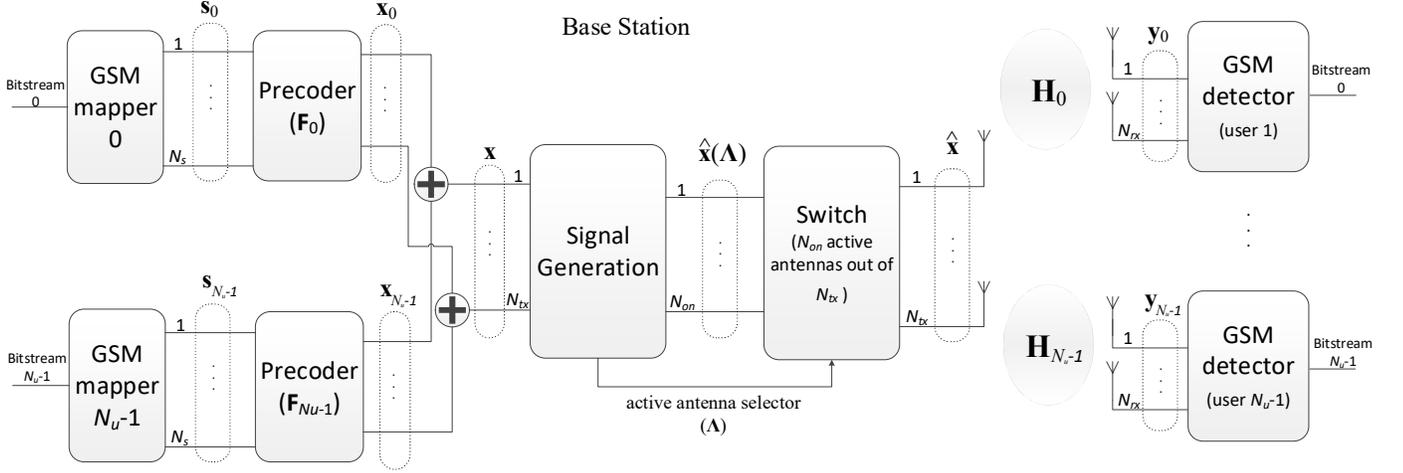

Fig. 1. Proposed PTSDIM system model.

and allows SU GSM detection. Instead of the usual approach where spatial information selects active receiver antennas [11]-[14], in PTSDIM spatial information is mapped onto the transmitter. This enables the system to work with large dimensional antenna settings which are much easier to deploy at the BS than at the users, thus achieving higher SE.

- To exploit the implicit sparsity of PTSDIM signals and the availability of large numbers of transmit antennas which was not fully addressed in previous related works [15]-[19], we propose the transmission of an equivalent signal with reduced power that produces the same received signal at all the receivers. The approach is transparent to the receivers and, therefore, does not require additional complexity.
- We also propose a signal construction method that allows the use of a reduced number of active antennas. This is an approach not exploited in previous works, simplifying the hardware implementation at the BS. This method is also transparent to the receivers.

*Notation:* $(\cdot)^T$ and $(\cdot)^H$ denote the transpose and conjugate transpose of a matrix/vector, $\|\cdot\|_p$ is the $\ell_p$-norm of a vector, $\|\cdot\|_0$ is its cardinality, supp($\mathbf{x}$) returns the support of $\mathbf{x}$, $\lfloor \cdot \rfloor$ is the floor function and $\mathbf{I}_n$ is the $n \times n$ identity matrix.

## II. SYSTEM MODEL

Let us consider the downlink of a MU-MIMO system where a BS equipped with $N_{tx}$ antennas transmits to $N_u$ users, each with $N_{rx}$ antennas, as shown in Fig. 1. We consider that the signal is represented as $\mathbf{s} = \begin{bmatrix} \mathbf{s}_0^T \ldots \mathbf{s}_{N_u-1}^T \end{bmatrix}^T$, where $\mathbf{s}_k \in \mathbb{C}^{N_s \times 1}$ is an IM symbol containing the information conveyed to user $k$. Assuming identical rates for all users, the length $N_s$ of $\mathbf{s}_k$ (which, before precoding, is identical to a GSM symbol) is constrained to $N_s \leq N_{tx}/N_u$. Based on the concept of GSM, only $N_a$ positions of $\mathbf{s}_k$ will be nonzero (active indexes) and carry modulation symbols. Each user signal vector can thus be written as $\mathbf{s}_k = [\ldots, 0, s_k^0, 0, \ldots, 0, s_k^{N_a-1}, 0, \ldots]^T$ where $s_k^j \in \mathcal{A}$ ($j=0,\ldots,N_a-1$), with $\mathcal{A}$ denoting an $M$-sized complex valued constellation set. Part of the information selects an active index combination (AIC) from a total of $N_{comb} = 2^{\lfloor \log_2 \binom{N_s}{N_a} \rfloor}$ available per user. The rest of the information is mapped onto $N_a$ complex-valued symbols which results in a total of

$$N_{bits} = N_u \left\lfloor \log_2 \binom{N_s}{N_a} \right\rfloor + N_u N_a \log_2 M \quad (1)$$

bits per PTSDIM symbol. We consider the availability of CSIT which is used for preprocessing the symbols with a linear precoder $\mathbf{F} = \begin{bmatrix} \mathbf{F}_0 \ldots \mathbf{F}_{N_u-1} \end{bmatrix}$, where $\mathbf{F}_k \in \mathbb{C}^{N_{tx} \times N_s}$ ($k=0,\ldots,N_u-1$). In this case the transmitted signal can be written as

$$\mathbf{x} = \sum_{k=0}^{N_u-1} \mathbf{x}_k = \sum_{k=0}^{N_u-1} \mathbf{F}_k \mathbf{s}_k. \quad (2)$$

After propagating through a flat fading channel, the baseband signal received at the $k^{th}$ user can be represented using

$$\mathbf{y}_k = \mathbf{H}_k \mathbf{x}_k + \mathbf{H}_k \sum_{\substack{j=0 \\ j \neq k}}^{N_u-1} \mathbf{x}_j + \mathbf{n}_k, \quad (3)$$

where $\mathbf{H}_k \in \mathbb{C}^{N_{rx} \times N_{tx}}$ is the channel matrix between the BS and user $k$ and $\mathbf{n}_k \in \mathbb{C}^{N_{rx} \times 1}$ is the noise vector containing independent zero-mean circularly symmetric Gaussian samples with covariance $2\sigma^2 \mathbf{I}_{N_{rx}}$. The second term in (3) is MUI which can be eliminated using a BD method [20]. In this case, the resulting $\mathbf{HF}$ matrix, with $\mathbf{H} = \begin{bmatrix} \mathbf{H}_0^T \ldots \mathbf{H}_{N_u-1}^T \end{bmatrix}^T$, becomes block diagonal. In PTSDIM, most of the positions of $\mathbf{s}_k$ are zero, with part of the information encoded in the equivalent channel impulse responses. Furthermore, due to the sparse nature of $\mathbf{s}_k$ it is possible to operate with a size $N_s \geq N_{rx}$ by resorting to compressed sensing based reconstruction techniques. This prevents the direct application of the power loading approaches presented in [20]. Therefore, we assume a simple BD precoder without any power loading optimization (even though one can be employed for accomplishing power control between users). Each precoder matrix $\mathbf{F}_k$ is designed in order to enforce that $\mathbf{H}_i \mathbf{F}_k = \mathbf{0}$ for all $i \neq k$. It can be directly obtained from the null space of matrix $\tilde{\mathbf{H}}_k = \begin{bmatrix} \mathbf{H}_0^T \ldots \mathbf{H}_{k-1}^T \mathbf{H}_{k+1}^T \ldots \mathbf{H}_{N_u-1}^T \end{bmatrix}^T$, which models the propagation of the signal intended for user $k$ to all the other receivers (MUI generated by the user). An orthonormal basis of the null space can be found from the singular value decomposition (SVD) given by

$$\tilde{\mathbf{H}}_k = \tilde{\mathbf{U}}_k \tilde{\mathbf{\Lambda}}_k \begin{bmatrix} \tilde{\mathbf{V}}_k^{(1)} & \tilde{\mathbf{V}}_k^{(0)} \end{bmatrix}^H, \quad (4)$$

where $\tilde{\mathbf{U}}_k$ contains the left-singular vectors, $\tilde{\mathbf{\Lambda}}_k$ is a rectangular diagonal matrix with the decreasing nonzero singular values, $\tilde{\mathbf{V}}_k^{(1)}$ comprises the right singular vectors corresponding to the nonzero singular values and $\tilde{\mathbf{V}}_k^{(0)}$ contains the remainder right singular vectors which span the null space of $\tilde{\mathbf{H}}_k$. If we set each precoder matrix as $\mathbf{F}_k = \tilde{\mathbf{V}}_k^{(0)}(:,1:N_s)$ then we know that $\tilde{\mathbf{H}}_k \mathbf{F}_k = \mathbf{0}$. This allows us to rewrite (3) as

$$\mathbf{y}_k = \mathbf{H}_k \mathbf{F}_k \mathbf{s}_k + \sum_{\substack{j=0 \\ j \neq k}}^{N_u-1} \mathbf{H}_k \mathbf{F}_j \mathbf{s}_j + \mathbf{n}_k = \hat{\mathbf{H}}_k \mathbf{s}_k + \mathbf{n}_k, \quad (5)$$

with $\hat{\mathbf{H}}_k = \mathbf{H}_k \mathbf{F}_k$, which shows that MUI has been removed. Since (5) is basically a SU GSM received signal model, detection can be accomplished using a conventional GSM receiver such as the ordered block minimum mean-squared error (OB-MMSE) detector [21]. It is important to note that while we have assumed a flat fading channel (which is reasonable in several scenarios or in orthogonal frequency division multiplexing systems), the proposed scheme can be directly extended to frequency selective channels by working with block based single-carrier transmissions. In this case, it is necessary to work with an enlarged channel matrix and signal vector in the received signal model (3), similarly to [22][23].

III. ENHANCED SIGNAL GENERATION

The direct application of the proposed BD precoder does not exploit the sparsity of $\mathbf{s}$. In this section we describe two different approaches for implementing the transmission. Both methods are completely transparent to the receivers, which do not need to know if the BS transmitted the original precoded signal or the alternative one. Denoting the active AIC for user $k$ as $I_k = \text{supp}(\mathbf{s}_k)$, we can write the noise-free received signal as

$$\mathbf{r}_k = \hat{\mathbf{H}}_k(:,I_k) \mathbf{s}_k(I_k), \quad (6)$$

where $\hat{\mathbf{H}}_k(:,I_k)$ contains only the $N_a$ columns of $\hat{\mathbf{H}}_k$ indexed by $I_k$ and $\mathbf{s}_k(I_k)$ is the reduced $N_a \times 1$ vector containing the nonzero elements of $\mathbf{s}_k$.

A. Transmitted Power Minimization

In the first approach, we design the transmitted signal with minimal power while enforcing the received signals arriving at the receivers to match the ones resulting from direct BD precoding at that time instance. Formally this can be written as

$$\min_{\hat{\mathbf{x}}} \|\hat{\mathbf{x}}\|_2^2 \quad \text{subject to} \quad \mathbf{r} = \mathbf{H}\hat{\mathbf{x}} \quad (7)$$

where $\mathbf{r} = \begin{bmatrix} \mathbf{r}_o^T \ldots \mathbf{r}_{N_u-1}^T \end{bmatrix}^T$. This formulation represents an equality constrained least-norm problem whose solution for a "fat" $\mathbf{H}$ (i.e., with $N_{tx} \geq N_{rx} N_u$) is simply

$$\hat{\mathbf{x}} = \mathbf{H}^H \left( \mathbf{H}\mathbf{H}^H \right)^{-1} \mathbf{r}. \quad (8)$$

B. Active Antenna Reduction

In the second approach, which we will refer to as active antenna reduction (AAR) method, the transmitted signal is redesigned in order to reduce the number of active antennas, $N_{on}$, while keeping the received signals close to the BD precoded ones. This problem can be formulated as

$$\min_{\hat{\mathbf{x}}} \|\mathbf{r} - \mathbf{H}\hat{\mathbf{x}}\|_2^2 \quad \text{subject to} \quad \|\hat{\mathbf{x}}\|_0 = N_{on}, \|\hat{\mathbf{x}}\|_2^2 \leq P_{\max} \quad (9)$$

with $P_{\max} = \|\mathbf{x}\|_2^2$, which is a regressor selection problem where the objective is to approximate vector $\mathbf{r}$ using a linear combination of a reduced number of columns of $\mathbf{H}$. To find the solution, we can first relax the cardinality constraint using the $\ell_1$-norm, as often done in cardinality based optimization problems in order to obtain convex formulations that can be solved efficiently [24]. We can then rewrite (9) as

$$\min_{\hat{\mathbf{x}}} \frac{1}{2}\|\mathbf{r} - \mathbf{H}\hat{\mathbf{x}}\|_2^2 + \lambda \|\hat{\mathbf{x}}\|_1 \quad \text{subject to} \quad \|\hat{\mathbf{x}}\|_2^2 \leq P_{\max} \quad (10)$$

where $\lambda > 0$ is a parameter commonly used in $\ell_1$-penalized least squares problems [25][26]. While a general interior point method can be applied to (10), the computational complexity is too high for the envisioned large problem settings. Therefore, we propose a proximal based iterative algorithm. First, we apply a projected gradient step [27] to the partial minimization problem

$$\min_{\tilde{\mathbf{x}} \in \mathcal{C}} \frac{1}{2}\|\mathbf{r} - \mathbf{H}\tilde{\mathbf{x}}\|_2^2 \triangleq f(\tilde{\mathbf{x}}), \quad (11)$$

which results in

$$\tilde{\mathbf{x}}^{(q)} = \Pi_{\mathcal{C}} \left( \hat{\mathbf{x}}^{(q-1)} - \gamma^{(q)} \mathbf{H}^H \left( \mathbf{H}\hat{\mathbf{x}}^{(q-1)} - \mathbf{r} \right) \right), \quad (12)$$

where $q$ is the iteration number, $\gamma$ is the step size, $\mathcal{C} = \left\{ \mathbf{x} \in \mathbb{C}^{N_{tx}} : \|\mathbf{x}\|_2^2 \leq P_{\max} \right\}$ and $\Pi_{\mathcal{C}}(\cdot)$ denotes the Euclidean projection onto $\mathcal{C}$. An estimate for the solution of the complete problem can then be obtained by finding an $\hat{\mathbf{x}}^{(q)}$ that minimizes the $\ell_1$-norm while still remaining close to the projected gradient step $\tilde{\mathbf{x}}^{(q)}$. This is equivalent to solving

$$\hat{\mathbf{x}}^{(q)} = \min_{\hat{\mathbf{x}}} \lambda \|\hat{\mathbf{x}}\|_1 + \frac{1}{2\gamma^{(q)}} \|\hat{\mathbf{x}} - \tilde{\mathbf{x}}^{(q)}\|_2^2, \quad (13)$$

which, by definition, corresponds to the proximal operator applied to $\tilde{\mathbf{x}}^{(q)}$, i.e., $\hat{\mathbf{x}}^{(q)} = \text{prox}_{\gamma^{(q)} \lambda \|\cdot\|_1}\left(\tilde{\mathbf{x}}^{(q)}\right)$ [28]. This proximal operator is given component-wise by the (complex) soft threshold function as $\hat{x}_i^{(q)} = \text{S}\left(\tilde{x}_i^{(q)}, \lambda \gamma^{(q)}\right)$ [29], with

**Algorithm 1:** Active Antenna Reduction (AAR)

1: **Input: r, H,** $\lambda$, $Q$
2: $\hat{\mathbf{x}}^{(0)} = \mathbf{z}^{(0)} = \breve{\mathbf{x}}^{(0)} = \breve{\mathbf{z}}^{(0)} = 0$, $\gamma = 1/\text{tr}(\mathbf{H}^H\mathbf{H})$
3: **for** $q=1,\ldots Q$ **do**
4: $\quad \tilde{\mathbf{x}}^{(q)} = \Pi_\mathcal{C}\left(\mathbf{z}^{(q-1)} - \gamma \mathbf{H}^H\left(\mathbf{H}\mathbf{z}^{(q-1)} - \mathbf{r}\right)\right)$
5: $\quad \hat{\mathbf{x}}^{(q)}(i) = \text{S}\left(\tilde{\mathbf{x}}^{(q)}(i), \lambda\gamma\right)$, $i = 0,\ldots,N_{tx}-1$
6: $\quad \mathbf{z}^{(q)} = \hat{\mathbf{x}}^{(q)} + \frac{q}{q+3}\left(\hat{\mathbf{x}}^{(q)} - \hat{\mathbf{x}}^{(q-1)}\right)$
7: **end for**
8: $\boldsymbol{\Lambda} \leftarrow \text{supp}\left(\hat{\mathbf{x}}^{(Q)}\right)$
9: **for** $q=1,\ldots Q$ **do** (polishing steps)
10: $\quad \breve{\mathbf{x}}^{(q)}(\boldsymbol{\Lambda}) = \Pi_\mathcal{C}\left(\breve{\mathbf{z}}^{(q-1)}(\boldsymbol{\Lambda}) - \gamma^{(q)}\mathbf{H}(:,\boldsymbol{\Lambda})^H\left(\mathbf{H}(:,\boldsymbol{\Lambda})\breve{\mathbf{z}}^{(q-1)}(\boldsymbol{\Lambda}) - \mathbf{r}\right)\right)$
11: $\quad \breve{\mathbf{z}}^{(q)} = \breve{\mathbf{x}}^{(q)} + \frac{q}{q+3}\left(\breve{\mathbf{x}}^{(q)} - \breve{\mathbf{x}}^{(q-1)}\right)$
12: **end for**
13: **Output:** $\hat{\mathbf{x}} \leftarrow \breve{\mathbf{x}}$.

$$\text{S}(u,v) = \frac{\max(|u|-v,0)}{\max(|u|-v,0)+v}u. \quad (14)$$

To improve the typical slow convergence rate of gradient based methods [30], we apply the acceleration described in [28]. In this case, projection (12) is applied not to the gradient based update computed on the previous point $\hat{\mathbf{x}}^{(q)}$ but rather to a linear combination of the previous two points, $\hat{\mathbf{x}}^{(q)}$ and $\hat{\mathbf{x}}^{(q-1)}$, namely $\tilde{\mathbf{x}}^{(q+1)} = \Pi_\mathcal{C}\left(\mathbf{z}^{(q)} - \gamma^{(q)}\nabla f\left(\mathbf{z}^{(q)}\right)\right)$ with

$$\mathbf{z}^{(q)} = \hat{\mathbf{x}}^{(q)} + \frac{q}{q+3}\left(\hat{\mathbf{x}}^{(q)} - \hat{\mathbf{x}}^{(q-1)}\right). \quad (15)$$

This results in (12) being replaced by

$$\tilde{\mathbf{x}}^{(q+1)} = \Pi_\mathcal{C}\left(\mathbf{z}^{(q)} - \gamma^{(q)}\mathbf{H}^H\left(\mathbf{H}\mathbf{z}^{(q)} - \mathbf{r}\right)\right). \quad (16)$$

Regarding the step size, we adopt a constant step given by $\gamma = 1/\text{tr}(\mathbf{H}^H\mathbf{H})$ (since $\|\mathbf{H}^H\mathbf{H}\|_2 \leq \text{tr}(\mathbf{H}^H\mathbf{H})$).

After applying the proposed $\ell_1$–based heuristic, one can try to find a better solution through an additional polishing step where (9) is solved with the sparsity pattern of $\hat{\mathbf{x}}$ fixed as $\boldsymbol{\Lambda} \leftarrow \text{supp}\left(\hat{\mathbf{x}}^{(Q)}\right)$ (obtained as the positions of the $N_{on}$ largest magnitude elements). In this case, a simple projected gradient is applied which consists on the following iteration step

$$\breve{\mathbf{x}}^{(q+1)}(\boldsymbol{\Lambda}) = \Pi_\mathcal{C}\left(\breve{\mathbf{x}}^{(q)}(\boldsymbol{\Lambda}) - \gamma^{(q)}\mathbf{H}(:,\boldsymbol{\Lambda})^H\left(\mathbf{H}(:,\boldsymbol{\Lambda})\breve{\mathbf{x}}^{(q)}(\boldsymbol{\Lambda}) - \mathbf{r}\right)\right). \quad (17)$$

Algorithm 1 summarizes all the steps, with $Q$ denoting the maximum number of iterations. It is important to note that we could have replaced the cardinality constraint in (9) with the $\ell_2$-norm as it is also convex and the corresponding proximal operator is simple to compute (as shown in [28]). However, contrary to the $\ell_1$-norm which induces sparseness of the solution, the $\ell_2$-norm tends to provide solutions where generally all components are nonzero, complicating the intended regressor selection [26].

*C. Computational Complexity*

The computation of the SVD in (4), has the heaviest cost with a complexity order of $O\left(N_u^4 N_{rx}^3 + N_u^2 N_{tx}^2 N_{rx}\right)$. Besides this, there is an additional cost resulting from the adoption of the two proposed methods. For the case of transmitted power minimization, the matrix inverse in (8) is the most expensive computation resulting in an overall complexity in terms of complex-valued floating-point operations (FLOPS) of

$$C_{\text{min.Power}} = N_u^3 N_{rx}^3 + N_u^2 N_{rx}^2\left(N_{tx} + 3/2\right) + \\ + N_u N_{rx}\left(3N_{tx} - 3/2\right) - N_{tx}. \quad (18)$$

In the AAR method, the most expensive step corresponds to the calculation of $\mathbf{H}^H\mathbf{H}$. In this case the total complexity is

$$C_{AAR} = N_{tx}^2\left(N_u N_{rx} - 1/2\right) + N_{tx}\left(4N_u N_{rx} - 3/2\right) + \\ + Q\left[2N_{tx}^2 + 10N_{tx} + 2N_{on}^2 + 6N_{on} - 2\right], \quad (19)$$

with the cost per iteration having an order of only $O(N_{tx}^2)$. It is important to note that the most expensive computations in both methods only have to be performed when $\mathbf{H}$ changes and, thus, in slowly varying channels the actual complexity will often be much smaller: $O\left(N_u^2 N_{rx}^2 + N_u N_{rx} N_{tx}\right)$ for the minimum power method and $O\left(N_{tx} N_u N_{rx} + Q N_{tx}^2\right)$ for AAR.

## IV. NUMERICAL RESULTS

In this section, we evaluate the performance of the proposed PTSDIM schemes. It is assumed that all users experience the same path-loss with all the elements of $\mathbf{H}$ being independently drawn according to a complex Gaussian distribution $\mathcal{CN}(0,1)$. Random modulated symbols are transmitted on the active positions with $E\left[\left|s_k^j\right|^2\right] = 1$. Fig. 2 shows the bit error rate (BER) as a function of the signal to noise ratio (SNR) per user for different configurations of PTSDIM in a MU scenario with $N_u$=15, $N_{tx}$=105, $N_s$=7, $N_a$=2, $N_{rx}$=4 and 64-QAM. This corresponds to a SE of 16 bits per channel use (bpcu) and per user. An OB-MMSE detector [21] is employed at each receiver. The PTSDIM with minimum transmission power achieves the best performance, with a gain of 11dB over the simple PTSDIM at a BER of $10^{-4}$. This was expected since the 'min. power' method produces the exact same received signal and, thus, the same BER as the simple PTSDIM. The difference is that it uses the minimum amount of transmitted power possible, being optimal in that sense (solution of the minimization problem (7)). The BER curve is basically a shifted version of the simple PTSDIM curve. Regarding the curves with AAR, we can observe that it is possible to improve the performance of the simple PTSDIM using fewer active antennas. In fact, with only $N_{on}$=80 (a reduction of 23%), the performance becomes very close to the minimum power curve. It is important to note that AAR is achieving this result with a higher complexity cost (4318830 versus 618105 flops according to (18) and (19)). In fact, while increasing $N_{on}$ could make the performance basically match the min. power curve, it would require a higher complexity cost and also sacrifice the main target of AAR which is to reduce the number of active antennas and respective RF chains. Another important aspect is that when the number of active antennas becomes too low, the algorithm may not be able to find an exact representation of the received signal **r** using a linear combination of $N_{on}$ columns of **H**. As the intended signal approximation degrades, it will cause the residual defined as $\mathbf{R} = \mathbf{r} - \mathbf{H}\hat{\mathbf{x}}$ to become nonzero and grow. This means that the received signal will contain an error even if no noise is present and this effect will manifest itself through an irreducible BER floor (case of $N_{on}$=60). It can also be observed that the number of iterations $Q$ adopted in the AAR algorithm can have a strong impact in its performance. Even though the cost per iteration is not high, it may be necessary to run a large number of iterations for the algorithm to converge to a solution. For the case of $N_{on}$=70, large performance gains are observed until $Q$=100. From then on, the improvements become smaller and the additional complexity cost may not be worthwhile. From the results, it can be observed that the use of acceleration provides a substantial performance improvement for the same number of iterations since the algorithm converges faster to a solution. Furthermore, it can also be seen that the polishing steps are critical for achieving a good performance since the first part of the algorithm basically aims at selecting the active antennas while the polishing part optimizes the signals to be transmitted from those.

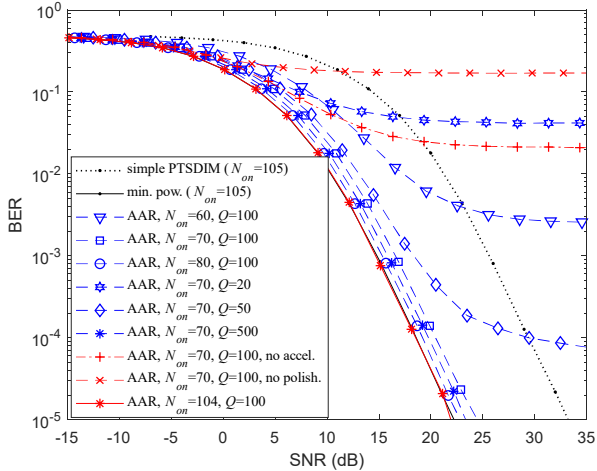

Fig. 2. BER performance of PTSDIM in a MU scenario with $N_u$=15, $N_{tx}$=105, $N_s$=7, $N_a$=2, $N_{rx}$=4, 64-QAM (16 bpcu per user).

Fig. 3 compares the BER performance of PTSDIM with AAR against the conventional BD MU-MIMO from [20], for three different SEs: 4, 6 and 8 bpcu per user. Since our proposed scheme can be directly extended to the reduced GSM version of [19] where all active antenna elements transmit the same symbol, we also included a curve for this type of (virtual) GSM, which we refer to as "reduced PTSDIM" (for 8 bpcu only, since the other two cases are in fact SM transmissions). The number of total active antennas in PTSDIM ($N_{on}$) is always kept the same as in BD MU-MIMO. It can be seen that achieving a higher SE through additional antennas while keeping the number of active antennas fixed and using a small modulation (PTSDIM with AAR), can be advantageous in terms of SNR when compared with the use of a higher modulation order (BD MU-MIMO and reduced PTSDIM). It is important to note, however, that while substantial gains can be obtained with the proposed approach, they will depend on the dimensions adopted for the PTSDIM. In fact, using a smaller length $N_s$ with $N_a$ fixed degrades the performance and can make PTSDIM underperform against BD MU-MIMO, as happens in the case of the SE of 4 bpcu. Regarding the complexity, the first steps of PTSDIM and BD MU-MIMO are both based on the SVD of matrices $\tilde{\mathbf{H}}_k$ and have the highest cost. Whereas the complexity order is the same, i.e., $O\left(N_u^4 N_{rx}^3 + N_u^2 N_{tx}^2 N_{rx}\right)$, $N_{tx}= N_{on}$ for BD MU-MIMO and $N_{tx}>N_{on}$ for PTSDIM with AAR, which means the later requires an additional complexity cost at the BS for achieving the presented gains.

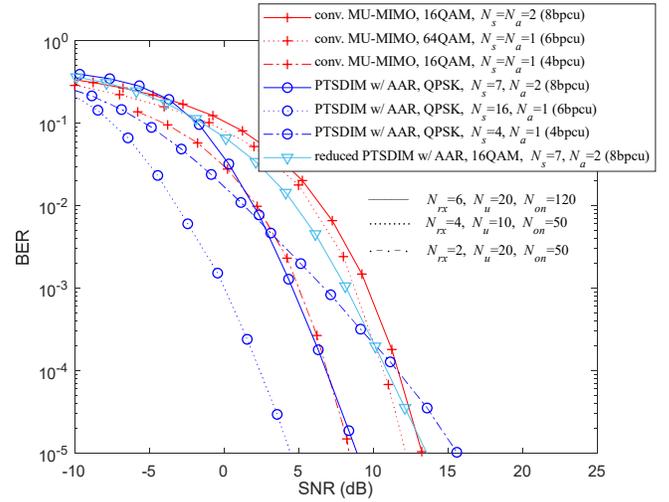

Fig. 3. BER performance of PTSDIM with AAR ($N_{tx}= N_s N_u$) and conventional BD MU-MIMO ($N_{tx}= N_{on}$) for different configurations.

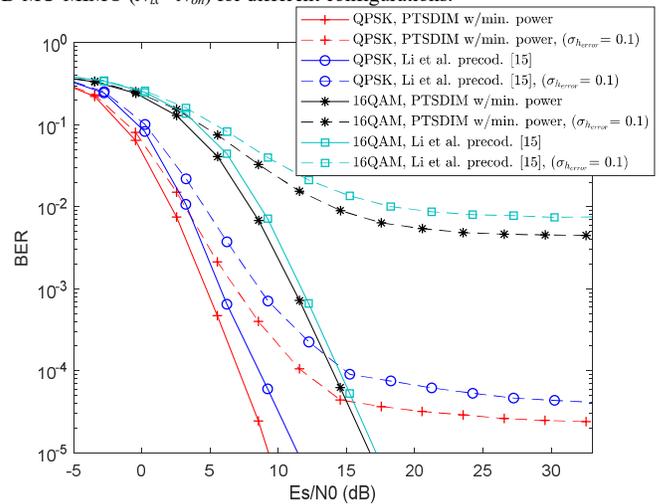

Fig. 4. BER performance of min. power PTSDIM and of the scheme from [15] for the scenario: $N_u$=12, $N_{tx}= N_{on}$=108, $N_s$=9, $N_a$=3, $N_{rx}$=6, with QPSK (12 bpcu per user) and 16-QAM (16 bpcu per user). Curves with perfect and imperfect CSIT ($\sigma_{h_{error}} = 0.1$) are shown.

In Fig. 4 we compare the performance of min. power PTSDIM against the precoder scheme proposed in [15] (designed for SM but can be directly extended to GSM). The two scenarios are based on the following configurations: $N_u$=12, $N_{tx}= N_{on}$=108, $N_s$=9, $N_a$=3, $N_{rx}$=6 with QPSK (12 bpcu per user) and also 16QAM (16 bpcu per user). An important aspect of GSMs and of any precoding scheme is the impact of imperfect channel knowledge in realistic scenarios [31][32]. Therefore, curves considering imperfect CSIT are also included where the assumed MIMO channel follows the same model used in [32]. In this case $\mathbf{H} = \bar{\mathbf{H}} + \mathbf{H}_{error}$, where $\bar{\mathbf{H}}$ is the CSI available at the transmitter and $\mathbf{H}_{error}$ denotes the CSI error. The entries of these two matrices are drawn according to complex Gaussian distributions of $\mathcal{CN}(0, \sigma_{\bar{h}}^2)$ and $\mathcal{CN}(0, \sigma_{h_{error}}^2)$, with $\sigma_{\bar{h}}^2 + \sigma_{h_{error}}^2 = 1$. For the curves shown in Fig. 4 we considered $\sigma_{h_{error}} = 0.1$ for the CSI error. Looking at the results, we can see that the proposed PTSDIM is able to perform better than the scheme from [15], with gains of 1.5 dB (QPSK) and 0.5 dB (16QAM) at a BER of $10^{-4}$, for the case of perfect CSIT. With imperfect CSIT, the performances of both schemes degrade, with the higher order modulation, 16-QAM, showing a higher sensitivity to channel estimation errors, as expected. Still, the proposed PTSDIM manages to achieve the best performance.

## V. Conclusions

In this paper we described a MU-MIMO system where a base station transmits precoded space domain IM symbols. The spatial bits are mapped at the transmitter side and combined with a precoder which allows the proposed MU-MIMO scheme to become equivalent to several independent SU GSM-MIMO transmissions. Two different signal construction methods were proposed enabling a more efficient system implementation, with potential performance gains over conventional precoded MU-MIMO.